\documentclass[manuscript]{aastex}

\shorttitle{Stellar Astrophysics with a dFTS. II.}
\shortauthors{Behr et al.}

\begin{document}

\title{Stellar Astrophysics with a \\
Dispersed Fourier Transform Spectrograph. \\
II. Orbits of Double-lined Spectroscopic Binaries}

\author{Bradford B. Behr\altaffilmark{1,2},
Andrew T. Cenko\altaffilmark{1}, 
Arsen R. Hajian\altaffilmark{1},
Robert S. McMillan\altaffilmark{3},
Marc Murison\altaffilmark{4},
Jeff Meade\altaffilmark{1}, and
Robert Hindsley\altaffilmark{5}}

\altaffiltext{1}{Department of Systems Design Engineering, University of Waterloo, Waterloo ON N2L 3G1, Canada.}
\altaffiltext{2}{Eureka Scientific, 2452 Delmer Street Suite 100, Oakland, CA 94602-3017.}
\altaffiltext{3}{Lunar and Planetary Laboratory, University of Arizona, Tucson, Arizona 85721.}
\altaffiltext{4}{US Naval Observatory, Flagstaff Station, 10391 W.\ Naval Observatory Rd., Flagstaff, AZ, 86001.}
\altaffiltext{5}{Remote Sensing Division, Naval Research Laboratory, Code 7215, Washington, DC, 20375.}


\begin{abstract}

We present orbital parameters for six double-lined spectroscopic binaries ($\iota$~Pegasi, $\omega$~Draconis, 12~Bo\"otis, V1143~Cygni, $\beta$~Aurigae, and Mizar~A) and two double-lined triple star systems ($\kappa$~Pegasi and $\eta$~Virginis). The orbital fits are based upon high-precision radial velocity observations made with a dispersed Fourier Transform Spectrograph, or dFTS, a new instrument which combines interferometric and dispersive elements. For some of the double-lined binaries with known inclination angles, the quality of our RV data permits us to determine the masses $M_1$ and $M_2$ of the stellar components with relative errors as small as 0.2\%.

\end{abstract}



\section{Introduction}

For the past several years, our research group has been developing a new optical spectrograph concept called the dispersed Fourier Transform Spectrograph, or dFTS. The instrument design merges a traditional Fourier Transform Spectrometer (FTS) with a dispersive grating spectrograph, such that the interferometer output is divided into thousands of narrowband channels, all operating in parallel. This multiplex advantage boosts the  effective throughput of the system by a large factor, making the dFTS competitive with echelle spectrographs for spectroscopic analysis of stars, particularly measurement of their radial velocities (RVs).

\citet{haj07} describes our prototype device, dFTS1, and explains the underlying theory and hardware implementation in detail. Based upon our commissioning observations with dFTS1, we subsequently designed and built a second-generation version, dFTS2, which we deployed to the Steward Observatory 2.3-meter Bok Telescope for a year-long observing campaign. In \citet{beh09}, we discuss the dFTS2 hardware and present velocimetry measurements of RV standard stars and single-lined spectroscopic binary stars (SB1s).

In this paper, we describe the results from our dFTS2 observations of double-lined spectroscopic binaries (SB2s) and double-lined triple systems. SB2s provide one of the best means for measuring the masses of stars: given an accurate RV curve for each stellar component and the inclination angle $i$ of the orbital plane to the observer's line of sight, we can derive the component masses using Kepler's Third Law. Traditional spectroscopic observations with an echelle spectrograph and thorium-argon calibration source can achieve velocity precision of $\sim 0.01$--0.10~km/s on late-type narrow-lined stars \citep{ram04, sku04, tom06, fek07, ram08}. For greater precision, \citet{kon05, kon09} has developed a technique using an iodine absorption cell, with which the RVs of a spectroscopic binary can be measured at the 0.005--0.010~km/s level. Our dFTS2 instrument, in contrast, achieves high RV precision and stability without a superposed reference spectrum. As described in \citet{beh09}, we measure the RVs of non-binary stars and single-lined binaries to 0.01--0.03~km/s, and anticipate even better performance once thermal stability issues in our instrument design have been addressed.


\section{Data acquisition and RV analysis procedure}

The data reported in this paper were collected between October 2007 and June 2008 during bright-time observing runs at the 2.3-meter Bok Telescope of the Steward Observatory on Kitt Peak. An observation of a given SB2 target consisted of 500 exposures spanning a range of interferometer delays corresponding to a spectral resolution of approximately 50,000. Each exposure lasted 1.0 to 4.0 seconds in duration, depending on the star's brightness and the atmospheric seeing and opacity. Each scan also required a total overhead time of approximately four minutes, independent of exposure time, for CCD readout and moving to the next delay position. Our targets, listed in Table~\ref{star-table}, were chosen because of their relative brightness and short periods, so that we could acquire many observations per star during the limited duration of this initial observing campaign.

\begin{deluxetable}{lccccl}
\tabletypesize{\scriptsize}
\tablewidth{0pt}
\tablecaption{Spectroscopic binary targets observed with dFTS2.\label{star-table}}
\tablehead{\colhead{star} & \colhead{$V$ magnitude} & \colhead{spectral type} & \colhead{$P$ (days)} & \colhead{$v \sin i_{A, B}$ (km/s)} & \colhead{references}}
\startdata
$\iota$ Peg			&3.8			&F5V			&10.21				&\phn 7.6, \phn 7.2		&\citet{fek83}							\\
$\omega$ Dra		&4.8			&F5V			&\phn 5.28		&\phn 7.1, \phn 6.6		&\citet{may87}, \citet{fek09}	\\
12 Boo					&4.8			&F8IV		&\phn 9.60		&13.1, 10.4					&\citet{bod05}, \citet{tom06}	\\
V1143 Cyg			&5.9			&F5V			&\phn 7.64		&23.6, 36.2					&\citet{and87}							\\
$\beta$ Aur			&1.9			&A2IV		&\phn 3.96		&34.5, 35.0					&\citet{smi48}, \citet{pou00}	\\
Mizar A					&2.3			&A2V		&20.54				&32.6, 36.2					&\citet{feh61}, \citet{pou00}		\\
$\kappa$ Peg		&4.2			&F5IV		&\phn 5.97		&\phn 7.1, 47.9			&\citet{may87}, \citet{haj07}		\\
$\eta$ Vir				&3.9			&A2V		&71.79				&\phn 5.1, \phn 4.5		&\citet{har92}, \citet{hum03}	\\
\enddata
\end{deluxetable}

To measure double-lined RVs from our interferogram data, we employed a variation of the standard two-dimensional cross-correlation technique \citep{maz94}. Instead of transforming our interferograms into spectra, and then cross-correlating model templates against each observed spectrum, we convert the template spectra into template interferograms, and then compare those model interferograms to the observed interferograms. Because interferograms add linearly, we can compute sequences of single-lined interferograms for the A and B components of a binary, spanning a range of radial velocities for each, and then add together A and B interferograms to create a two-dimensional grid of double-lined model interferograms. Calculating the $\chi^2$ difference between the models and the observed interferogram data, we construct a map of fit quality, where the minimum point indicates the best-fit solution for component velocities $v_1$ and $v_2$, and the projection of the $\chi^2 = \chi^2_{\rm min} + 2.30$ contour line onto the $v_1$ and $v_2$ axes provides 1-$\sigma$ error bars on each radial velocity point. The shapes of the $\chi^2$ contours were elliptical with minor and major axes aligned to the $v_1$ and $v_2$ axes, indicating no significant covariance between the component velocity errors.

For optimal results, the template spectra must be well matched to the actual spectra of the two stellar components. We generated synthetic spectra using the SPECTRUM spectral synthesis package \citep{gra94} (see also \url{http://www.phys.appstate.edu/spectrum/spectrum.html}), and then varied the transition strength $\log g\!f$ of each line and the projected rotation velocity $v \sin i$ for each stellar component to minimize the $\chi^2$ difference between the model and the observed data. A final template spectrum for each component was calculated from a median-filtered average of transition strengths from the individual observations, and the final two-dimensional RV cross-correlation was then performed using these templates. We found that we could derive relatively precise and self-consistent $v \sin i$ values from our interferograms (as listed in Table~\ref{star-table}); a future paper will explore the use of dFTS data to measure stellar rotation velocities and other line broadening mechanisms. 

\begin{deluxetable}{lcccccc}
\tabletypesize{\scriptsize}
\tablewidth{0pt}
\tablecaption{Radial velocity data measured with dFTS2.\label{rv-table}}
\tablehead{\colhead{star} & \colhead{${\rm HJD}-2,400,000$} & \colhead{phase} & \colhead{$V_1$ (km~s${}^{-1}$)} & \colhead{$V_1$ error (km~s${}^{-1}$)} & \colhead{$V_2$ (km~s${}^{-1}$)} & \colhead{$V_2$ error (km~s${}^{-1}$)}}
\startdata
$\iota$ Peg  &54,400.6388  &0.109  &33.302  &0.017  &-64.348  &0.080	\\
$\iota$ Peg  &54,401.7122  &0.214  &6.584  &0.019  &-21.902  &0.091	\\
$\iota$ Peg  &54,401.7269  &0.215  &6.235  &0.017  &-20.958  &0.084	\\
$\iota$ Peg  &54,402.6529  &0.306  &-20.809  &0.021  &22.781  &0.109	\\
$\iota$ Peg  &54,403.6604  &0.404  &-44.161  &0.020  &59.933  &0.103	\\
$\iota$ Peg  &54,404.6147  &0.498  &-52.709  &0.032  &73.279  &0.196	\\
$\iota$ Peg  &54,604.9660  &0.115  &31.993  &0.017  &-62.573  &0.080	\\
$\iota$ Peg  &54,606.9635  &0.311  &-22.230  &0.015  &24.831  &0.067	\\
$\iota$ Peg  &54,634.8758  &0.044  &42.325  &0.015  &-78.934  &0.067	\\
$\iota$ Peg  &54,634.8959  &0.046  &42.119  &0.016  &-78.687  &0.072	\\
$\iota$ Peg  &54,635.9132  &0.145  &25.341  &0.017  &-51.490  &0.076	\\
$\iota$ Peg  &54,635.9342  &0.147  &24.823  &0.016  &-50.598  &0.075	\\
$\iota$ Peg  &54,637.8866  &0.339  &-29.723  &0.017  &36.656  &0.073	\\
$\iota$ Peg  &54,637.9068  &0.341  &-30.242  &0.017  &37.434  &0.080	\\
$\iota$ Peg  &54,638.9426  &0.442  &-49.514  &0.015  &68.427  &0.066	\\
$\iota$ Peg  &54,638.9628  &0.444  &-49.713  &0.015  &68.752  &0.064	\\
\hline
$\omega$ Dra  &54,547.9814  &0.160  &-48.332  &0.027  &29.389  &0.049	\\
$\omega$ Dra  &54,549.9958  &0.542  &19.025  &0.026  &-53.678  &0.046	\\
$\omega$ Dra  &54,578.9388  &0.024  &-44.148  &0.031  &24.305  &0.056	\\
$\omega$ Dra  &54,579.9848  &0.222  &-41.842  &0.027  &21.495  &0.050	\\
$\omega$ Dra  &54,603.8020  &0.733  &13.203  &0.024  &-46.284  &0.045	\\
$\omega$ Dra  &54,603.9267  &0.756  &9.247  &0.029  &-41.483  &0.054	\\
$\omega$ Dra  &54,604.9389  &0.948  &-31.906  &0.026  &9.346  &0.049	\\
$\omega$ Dra  &54,605.8126  &0.114  &-49.790  &0.023  &31.110  &0.041	\\
$\omega$ Dra  &54,605.9378  &0.137  &-49.461  &0.026  &30.731  &0.048	\\
$\omega$ Dra  &54,634.9202  &0.627  &22.606  &0.025  &-57.902  &0.046	\\
$\omega$ Dra  &54,634.9758  &0.637  &22.298  &0.023  &-57.528  &0.040	\\
$\omega$ Dra  &54,635.6924  &0.773  &6.250  &0.022  &-37.884  &0.039	\\
$\omega$ Dra  &54,636.6824  &0.960  &-34.305  &0.024  &12.288  &0.042	\\
$\omega$ Dra  &54,636.9410  &0.009  &-42.260  &0.022  &22.042  &0.039	\\
$\omega$ Dra  &54,637.6927  &0.152  &-48.736  &0.027  &29.976  &0.049	\\
$\omega$ Dra  &54,637.9312  &0.197  &-45.017  &0.026  &25.330  &0.047	\\
\hline
12 Boo  &54,547.9153  &0.591  &-29.048  &0.065  &50.273  &0.072	\\
12 Boo  &54,548.9127  &0.694  &-48.375  &0.076  &70.324  &0.083	\\
12 Boo  &54,549.9111  &0.798  &-51.801  &0.080  &73.760  &0.093	\\
12 Boo  &54,577.7996  &0.702  &-49.395  &0.049  &71.239  &0.056	\\
12 Boo  &54,579.8185  &0.912  &-18.824  &0.060  &39.831  &0.070	\\
12 Boo  &54,579.8722  &0.918  &-16.030  &0.054  &37.061  &0.062	\\
12 Boo  &54,602.8623  &0.312  &50.008  &0.070  &-31.349  &0.079	\\
12 Boo  &54,603.7304  &0.402  &22.867  &0.073  &-3.201  &0.084	\\
12 Boo  &54,604.7012  &0.503  &-6.315  &0.061  &26.942  &0.071	\\
12 Boo  &54,605.7012  &0.607  &-32.713  &0.055  &53.903  &0.065	\\
12 Boo  &54,606.6579  &0.707  &-49.876  &0.063  &71.703  &0.071	\\
12 Boo  &54,635.7226  &0.733  &-52.290  &0.064  &74.099  &0.073	\\
12 Boo  &54,638.7263  &0.046  &58.158  &0.084  &-39.386  &0.092	\\
\hline
V1143 Cyg  &54,603.8512  &0.742  &6.959  &0.260  &-39.988  &0.490	\\
V1143 Cyg  &54,604.8585  &0.873  &57.691  &0.200  &-92.635  &0.333	\\
V1143 Cyg  &54,605.8830  &0.007  &59.725  &0.191  &-95.517  &0.309	\\
V1143 Cyg  &54,606.8736  &0.137  &-69.958  &0.250  &37.136  &0.445	\\
V1143 Cyg  &54,634.8268  &0.796  &22.287  &0.168  &-55.975  &0.320	\\
V1143 Cyg  &54,636.8467  &0.060  &-29.744  &0.205  &-0.671  &0.339	\\
V1143 Cyg  &54,638.8573  &0.323  &-65.668  &0.192  &33.420  &0.335	\\
\hline
$\beta$ Aur  &54,400.9449  &0.134  &54.375  &0.151  &-91.605  &0.149	\\
$\beta$ Aur  &54,401.9169  &0.379  &-96.561  &0.158  &63.013  &0.156	\\
$\beta$ Aur  &54,402.9819  &0.648  &-81.719  &0.193  &48.833  &0.195	\\
$\beta$ Aur  &54,402.9909  &0.651  &-81.088  &0.169  &47.313  &0.170	\\
$\beta$ Aur  &54,403.8922  &0.878  &60.619  &0.289  &-98.087  &0.286	\\
$\beta$ Aur  &54,403.8982  &0.880  &61.415  &0.203  &-98.274  &0.196	\\
$\beta$ Aur  &54,404.8387  &0.117  &62.259  &0.292  &-99.681  &0.290	\\
$\beta$ Aur  &54,404.8542  &0.121  &60.690  &0.241  &-97.396  &0.236	\\
$\beta$ Aur  &54,404.8680  &0.125  &58.476  &0.253  &-95.754  &0.244	\\
$\beta$ Aur  &54,487.7015  &0.042  &86.798  &0.292  &-124.300  &0.274	\\
$\beta$ Aur  &54,487.7106  &0.044  &86.270  &0.237  &-124.136  &0.236	\\
$\beta$ Aur  &54,487.7191  &0.046  &86.305  &0.381  &-123.834  &0.374	\\
$\beta$ Aur  &54,487.7530  &0.055  &84.279  &0.550  &-122.404  &0.556	\\
$\beta$ Aur  &54,488.7643  &0.310  &-56.848  &0.217  &23.889  &0.209	\\
$\beta$ Aur  &54,488.7731  &0.312  &-58.307  &0.206  &24.076  &0.203	\\
$\beta$ Aur  &54,488.7818  &0.315  &-59.895  &0.200  &26.191  &0.196	\\
$\beta$ Aur  &54,491.6686  &0.044  &85.997  &0.288  &-124.270  &0.283	\\
$\beta$ Aur  &54,491.6835  &0.047  &85.742  &0.159  &-123.753  &0.159	\\
$\beta$ Aur  &54,491.7588  &0.066  &81.639  &0.350  &-119.413  &0.341	\\
$\beta$ Aur  &54,491.7881  &0.074  &79.167  &0.413  &-116.398  &0.395	\\
\hline
Mizar  &54,488.9333  &0.660  &21.818  &0.361  &-37.796  &0.408	\\
Mizar  &54,488.9434  &0.661  &22.811  &0.980  &-37.749  &1.093	\\
Mizar  &54,488.9518  &0.661  &22.457  &0.312  &-37.301  &0.394	\\
Mizar  &54,488.9604  &0.662  &23.821  &0.556  &-37.302  &0.708	\\
Mizar  &54,492.0078  &0.810  &44.126  &0.247  &-57.208  &0.329	\\
Mizar  &54,492.0468  &0.812  &43.908  &0.354  &-57.704  &0.456	\\
Mizar  &54,548.8658  &0.578  &11.679  &0.482  &-26.290  &0.640	\\
Mizar  &54,549.8253  &0.625  &17.287  &0.174  &-32.974  &0.222	\\
Mizar  &54,576.7827  &0.937  &39.127  &0.417  &-52.615  &0.543	\\
Mizar  &54,578.7569  &0.034  &-74.826  &0.211  &59.789  &0.266	\\
Mizar  &54,579.7472  &0.082  &-81.159  &0.264  &66.865  &0.333	\\
Mizar  &54,579.7558  &0.082  &-81.404  &0.289  &66.645  &0.366	\\
Mizar  &54,602.7619  &0.202  &-52.427  &0.283  &37.381  &0.368	\\
Mizar  &54,603.6351  &0.245  &-42.830  &0.240  &27.669  &0.301	\\
Mizar  &54,604.6286  &0.293  &-33.460  &0.289  &19.960  &0.377	\\
Mizar  &54,634.6759  &0.756  &36.381  &0.306  &-49.740  &0.402	\\
Mizar  &54,636.6348  &0.851  &48.098  &0.184  &-61.666  &0.246	\\
Mizar  &54,637.6408  &0.900  &48.821  &0.213  &-62.867  &0.287	\\
\hline
$\kappa$ Peg  &54,400.6227  &0.069  &26.262  &0.031  &2.603  &0.361	\\
$\kappa$ Peg  &54,401.6788  &0.245  &-10.951  &0.035  &-0.613  &0.345	\\
$\kappa$ Peg  &54,402.6158  &0.402  &-46.681  &0.055  &4.407  &0.415	\\
$\kappa$ Peg  &54,402.6305  &0.405  &-47.035  &0.064  &3.843  &0.465	\\
$\kappa$ Peg  &54,403.6430  &0.574  &-49.989  &0.039  &0.649  &0.309	\\
$\kappa$ Peg  &54,404.5841  &0.732  &-16.839  &0.042  &-0.683  &0.416	\\
$\kappa$ Peg  &54,404.5988  &0.734  &-16.173  &0.054  &0.520  &0.506	\\
$\kappa$ Peg  &54,634.8531  &0.294  &-24.846  &0.029  &0.465  &0.317	\\
$\kappa$ Peg  &54,635.8924  &0.468  &-55.088  &0.032  &-0.740  &0.284	\\
$\kappa$ Peg  &54,636.8741  &0.632  &-41.896  &0.035  &6.552  &0.275	\\
$\kappa$ Peg  &54,637.8349  &0.793  &-2.007  &0.034  &0.883  &0.333	\\
\hline
$\eta$ Vir	  &54,491.9611  &0.229  &11.665  &0.022  &-11.833  &0.067	\\
$\eta$ Vir	  &54,548.8113  &0.021  &-27.988  &0.034  &40.160  &0.101	\\
$\eta$ Vir	  &54,549.7519  &0.034  &-26.057  &0.030  &37.392  &0.090	\\
$\eta$ Vir	  &54,576.7011  &0.409  &21.091  &0.035  &-25.264  &0.099	\\
$\eta$ Vir	  &54,577.7526  &0.424  &21.138  &0.035  &-25.293  &0.105	\\
$\eta$ Vir	  &54,579.7025  &0.451  &20.923  &0.039  &-25.080  &0.119	\\
$\eta$ Vir	  &54,603.7043  &0.785  &-6.225  &0.028  &10.650  &0.085	\\
$\eta$ Vir	  &54,604.6471  &0.799  &-8.312  &0.029  &13.633  &0.087	\\
$\eta$ Vir	  &54,605.6656  &0.813  &-10.717  &0.030  &16.643  &0.091	\\
$\eta$ Vir	  &54,634.6939  &0.217  &9.759  &0.033  &-10.524  &0.101	\\
$\eta$ Vir	  &54,636.6521  &0.244  &12.931  &0.026  &-14.812  &0.076	\\
\enddata
\end{deluxetable}

Because our template spectra are generated using atomic transition wavelengths from the NIST catalog, the derived RVs can be considered ``accurate'' in the sense that they reflect the total Doppler shift between the rest wavelength of a line and the observed wavelength. We have not (yet) attempted to tie our velocity scale to any IAU velocity standards, nor do we make any correction for gravitational redshift effects. The only adjustment made to the RV data is conversion to a solar system barycenter reference frame, using the IRAF tool {\tt bcvcorr}. These barycentric RV data are listed in Table~\ref{rv-table}. It should be noted that for the $\kappa$ Peg RVs, ``$V_1$'' refers to the Bb component and ``$V_2$'' refers to the A component.

We derived orbital parameters from our RV data points using the IDL routines {\tt CURVEFIT}, a gradient-expansion nonlinear least-squares fitting algorithm included with the IDL package, and {\tt HELIO\_RV} \citep{lan93}, which computes a line-of-sight velocity curve for a binary component given the period $P$, periastron time $T$ (or for circular orbits, the time of maximum positive velocity), eccentricity $e$, periastron longitude $\omega$, RV semi-amplitude $K$, and systemic velocity $V_0$ (alternatively denoted as $\gamma$ by some researchers). We fit the primary and secondary RV points simultaneously, assuming that $P$, $T$, $e$, and $V_0$ are the same for both components, and that $\omega_1$ and $\omega_2$ differ by $180^\circ$. The {\tt CURVEFIT} routine returns $1\sigma$ uncertainties (standard deviations) for all derived parameters. For all of our SB2 targets, we adopted the orbital period $P$ from previously-published analyses, because our observations covered a relatively short period of time. We did not correct for the light travel time across each binary system, because the resulting changes in the RV values are small compared to the RV error bars in all six cases.


\section{Results on double-lined binary systems}

\subsection{$\iota$ Pegasi}

\begin{figure}[ht]
\centering
\includegraphics[scale=0.7]{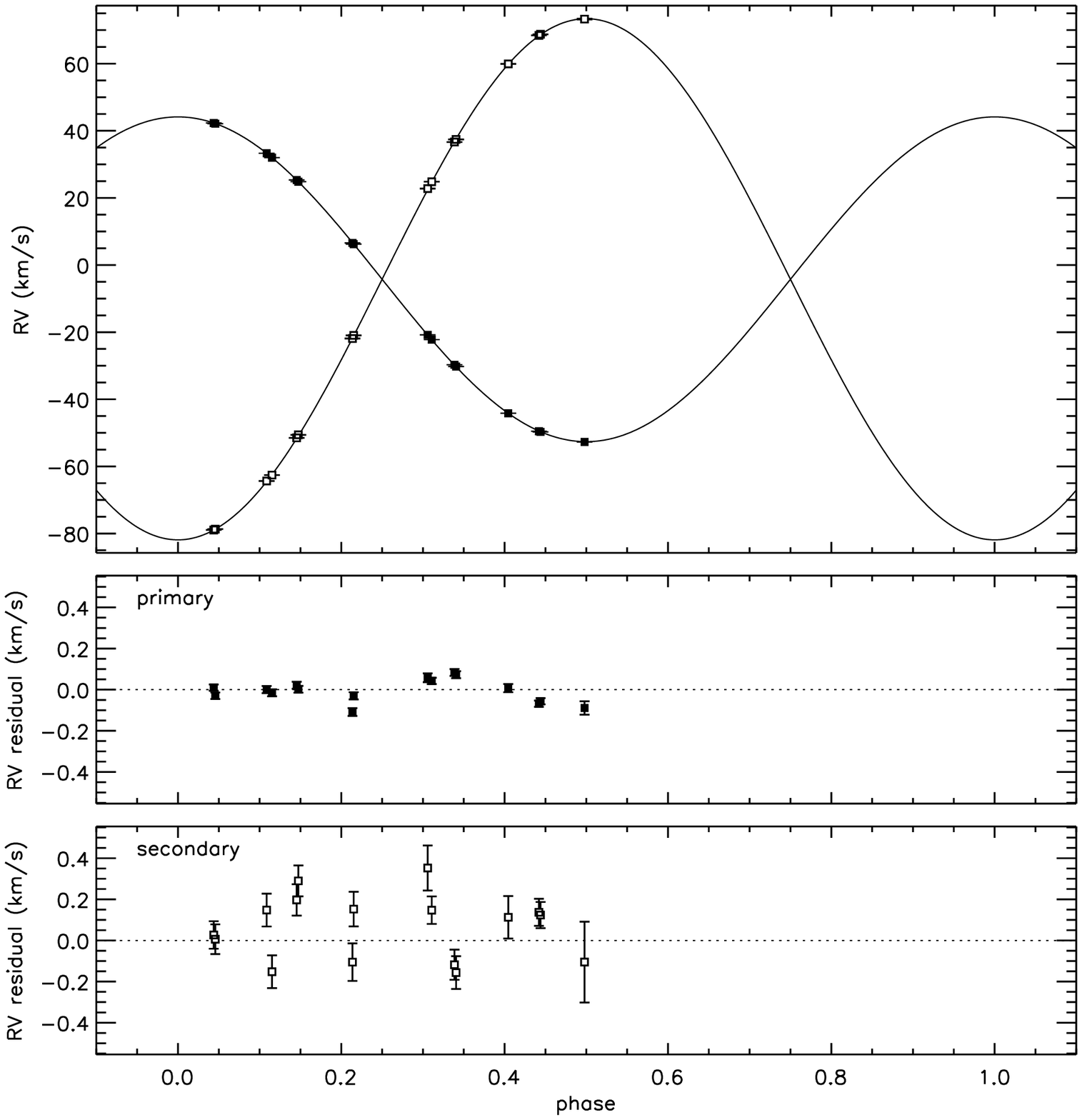}
\caption{Radial velocity measurements of the SB2 system $\iota$ Pegasi. Filled squares show the measured RV of the primary component, and open squares indicate the secondary component. Our observing campaign ended before we were able to complete the phase coverage of this system.\label{iotpeg-fig}}
\end{figure}

Our RV measurements for the double-lined spectroscopic binary $\iota$ Pegasi (HR 8430, HD 210027, HIP 109176) are plotted in Figure~\ref{iotpeg-fig}. The most recent published RV work on this system comes from \citet{fek83}, whose orbital parameters are listed in Table~\ref{iotpeg-table} along with the values that we derive from our dFTS2 observations. In addition to adopting their value of the system's orbital period, we also followed their lead in assuming a circular orbit, because our RV points only covered half of the orbital phase, and {\tt CURVEFIT} could not place meaningful constraints on $e$ or $\omega$. Our values for $K_1$ and $K_2$ are compatible with those of Fekel \& Tomkin, although our solution for $V_0$ differs by a statistically significant amount. This discrepancy may indicate the gravitational influence of a unseen and distant third stellar component of the system, although \citet{tok06} did not find any close tertiary companions in 2MASS images of $\iota$~Peg, and the astrometric observations of \citet{bod99} saw no evidence for a compansion either. Alternatively, the difference in $V_0$ might merely be a result of different RV zero points between Fekel \& Tomkin's observations and ours --- unfortunately, we did not make any RV observations of $\iota$~Piscium, the RV standard star that they used as their reference spectrum.

Although the RMS scatter of our RV points around the best-fit orbital curves is small (56 m/s and 158 m/s for the A and B components, respectively), the scatter is larger than would be expected from the error bars on each individual RV measurement, and is significantly above the instrumental RV error floor of $\sim 10$~m/s that we determined in \citet{beh09}. To account for this discrepancy, we multiply the RV error bars by 3.11 for the primary and 1.83 for the secondary, such that the mean per-measurement error bar matches the RMS deviation $\sigma_{\rm RV}$ for each stellar component. The orbital fits are then recalculated using these scaled error bars, and the resulting orbital parameters are listed in the rightmost column of Table~\ref{iotpeg-table}. This same procedure is applied to all subsequent binary systems as well.

The additional RV variability, if real, may be a result of stellar activity on both the primary and secondary, driven by tidal interactions between the two stars. Fekel \& Tomkin measure $v \sin i = 7 \pm 2$~km/s for the primary, very close to an estimated synchronous rate of 6.5~km/s, and $v \sin i = 9 \pm 3$~km/s for the secondary, which is well above the estimated synchronous rate of 4.5 km/s. \citet{gra84} finds $v \sin i$ values of $6.5 \pm 0.3$~km/s (primary) and $5 \pm 1$~km/s (secondary), suggesting that the system {\it is} synchronized. Our preliminary analysis of line broadening indicates $v \sin i = 7.6$ and 7.2 for the primary and secondary, respectively. If the secondary is indeed spinning more rapidly than the synchronous rate, then above-average surface activity could result, which would add significant astrophysical RV ``jitter'' to our measurements. A synchronously-rotating component would be less susceptible to tidal effects, but activity might still be enhanced by the proximity of a massive companion. However, \citet{kon09b} measured RVs of $\iota$~Peg with three different spectrographs, and found no jitter greater than 17 m/s (primary) and 85 m/s (secondary), suggesting that the jitter observed by dFTS2 was instrumental rather than astrophysical.

\citet{bod99} measured an inclination angle for this system of $i = 95.67^\circ \pm 0.22^\circ$ (based on their primary data set). Using this value along with our orbital parameters, with the fundamental parameters recommended by \citet{tor10}, we derive stellar masses of $M_1 = 1.3241 \pm 0.0018\ M_\odot$ and $M_2 = 0.8251 \pm 0.0010\ M_\odot$, which represent relative (statistical) errors of 0.14\% and 0.12\% respectively. With scaled error bars, the mass estimates are the same, albeit with larger error bars, for relative uncertainties of 0.19\% and 0.15\%. These values agree reasonably well with the calculations of Boden et al., who used Fekel \& Tomkin's $K_1$ and $K_2$ values to determine $M_1 = 1.326 \pm 0.016\ M_\odot$ and $M_2 = 0.819 \pm 0.009\ M_\odot$. For our mass estimates, the largest component of the error budget is due to the uncertainty in $i$, although the uncertainties in the $K$ values are also significant contributors.

\begin{deluxetable}{lccc}
\tabletypesize{\footnotesize}
\tablewidth{0pt}
\tablecaption{Orbital parameters and stellar mass estimates for $\iota$~Pegasi.\label{iotpeg-table}}
\tablehead{\colhead{parameter} & \colhead{\citet{fek83}} & \colhead{this work (formal RV errors)} & \colhead{this work (scaled RV errors)}}
\startdata
$P$ (days)						&$10.213033 \pm 0.000013$		&adopted from F\&T						&adopted from F\&T					\\
$T$ (reduced HJD)		&45320.1423								&$54399.5296 \pm 0.0003$		&$54399.5288 \pm 0.0007$	\\
$e$								&\phn 0.0 assumed						&\phn 0.0	 assumed						&\phn 0.0	 assumed					\\
$K_1$ (km/s)				&$48.1 \pm 0.2$							&$48.380 \pm 0.006$					&$48.380 \pm 0.018$				\\ 
$K_2$ (km/s)				&$77.9 \pm 0.3$							&$77.637 \pm 0.027$					&$77.638 \pm 0.050$				\\ 
$V_0$ (km/s)				&$-5.5 \pm 0.2\;$						&$-4.245 \pm 0.007\;$				&$-4.229 \pm 0.015\;$			\\
$N_{\rm obs}$				&32												&16												&16				\\
$\chi^2$ primary			&\nodata										&142.20										&16.45			\\
$\chi^2$ secondary		&\nodata										&64.52											&16.11			\\
$\sigma_{\rm RV}$ primary (km/s)		&0.90	 				&0.056											&0.059			\\
$\sigma_{\rm RV}$ secondary (km/s)	&1.16 					&0.158											&0.156			\\ \\
$M_1\ (M_\odot)$		&$1.326 \pm 0.016$~\tablenotemark{a}		&$1.3239 \pm 0.0018$				&$1.3239 \pm 0.0025$			\\		
$M_2\ (M_\odot)$		&$0.819 \pm 0.009$~\tablenotemark{a}		&$0.8250 \pm 0.0010$				&$0.8250 \pm 0.0013$			\\		
\enddata
\tablenotetext{a}{\ from \citet{bod99}, using F\&T velocities in conjunction with spatial interferometer observations}
\end{deluxetable}



\subsection{$\omega$ Draconis}

The spectroscopic orbit of the $\omega$ Draconis system (HR 6596, HD 160922, HIP 86201) was measured by \citet{may87} and more recently by \citet{fek09}. Their derived orbital parameters are shown in Table~\ref{omedra-table}, along with our values. Our $K$ velocities agree closely with those of Fekel et al. We find a small but nonzero eccentricity for the orbits, $0.0023 \pm 0.0002$. Fekel et al. derived $e = 0.0027 \pm 0.0008$ for the primary, with $\omega_1 = 40.1^\circ \pm 17.8^\circ$ \citep{fek09b}, but their $e$ and $\omega$ values for the secondary did not agree with those of the primary, so they adopted a circular orbit for the system. (Our $\omega_1 = 137.86 \pm 13.48$, as described below.) Despite the similarity between our $e$ value and their nonzero $e$ value, the measurements of $\omega_1$ are substantially different, so we cannot plausibly claim that an orbital eccentricity has been clearly detected. 

\begin{figure}[ht]
\centering
\includegraphics[scale=0.7]{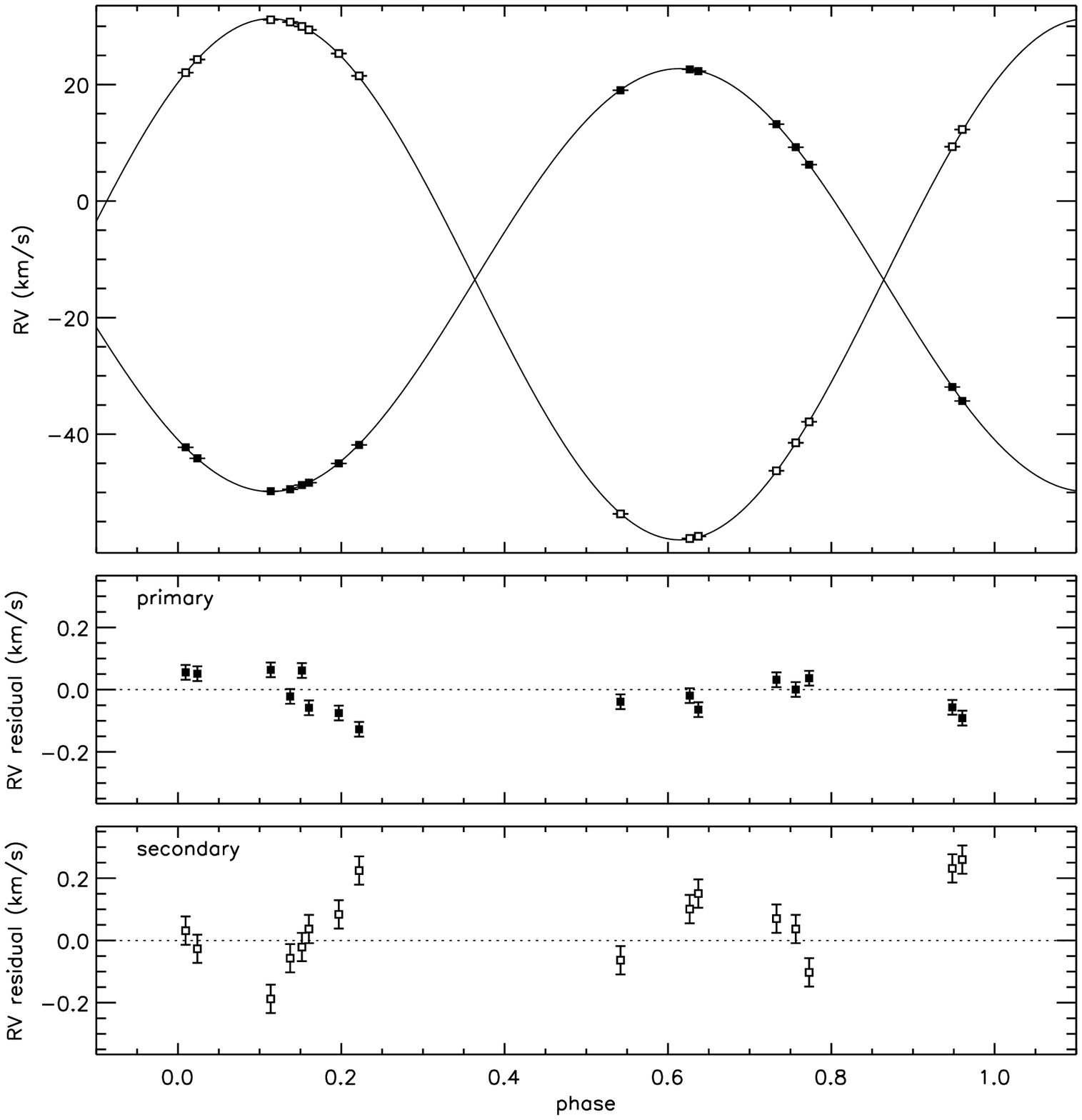}
\caption{Radial velocity measurements of the SB2 system $\omega$ Draconis.\label{omedra-fig}}
\end{figure}

As an additional test of the nonzero eccentricity, we follow the Fekel et al. procedure of fitting orbits to the A and B components separately and comparing the derived values for $\omega_1$ and $\omega_2$, which should differ by $180^\circ$. Using the formal error bar data, we calculated $e_1 = 0.0020 \pm 0.0003$ with $\omega_1 = 74.16^\circ \pm 12.74^\circ$, and $e_2 = 0.0027 \pm 0.0005$ with $\omega_2 = 243.14^\circ \pm 13.05^\circ$. The eccentricity values agree reasonably well, and the $\omega$ angles differ by $\sim 169^\circ$, which is within $1\sigma$ of $180^\circ$. However, this $\omega_1$ value does not agree with the $\omega_1$ value from the combined fit, which is puzzling. This discrepancy may be related to the apparent systematic trends in the secondary RV residuals which are evident in Figure~\ref{omedra-fig}. Further observations with better phase coverage will be required to validate or refute our measured eccentricity for $\omega$~Dra. 

As with $\iota$~Peg, our value for $V_0$ differs from the prior work by a statistically significant amount, although the magnitude of the difference is not as large. Differences in the RV zero point are the most likely explanation. 

No visual orbit or estimate of the inclination angle $i$ has been determined for this binary, despite efforts to resolve it using speckle interferometry \citep{iso91, iso95}. We are therefore unable to calculate the true masses of the stellar components. We find $M_1 \sin^3 i = 0.16054 \pm 0.00011\ M_\odot$ and $M_2 \sin^3 i = 0.13030 \pm 0.00007\ M_\odot$, in moderately good agreement with Fekel et al. (With the scaled RV error bars, our mass estimates are virtually unchanged, with error bars approximately three times larger.) We hope that long-baseline interferometers will soon be able to resolve the astrometric orbit of this system and determine the inclination angle.

\begin{deluxetable}{lcccc}
\tabletypesize{\scriptsize}
\tablewidth{0pt}
\tablecaption{Orbital parameters and stellar mass estimates for $\omega$ Draconis.\label{omedra-table}}
\tablehead{\colhead{parameter} & \colhead{\citet{may87}} & \colhead{\citet{fek09}} & \colhead{this work (formal RV errors)} & \colhead{this work (scaled RV errors)}}
\startdata
$P$ (days)						&$5.279799 \pm 0.000003$		&$5.2798088 \pm 0.0000083$ 		&adopted from Fekel et al.		&adopted from Fekel et al.			\\
$T$ (reduced HJD)		&$44698.273 \pm 0.005$			&$53980.1606 \pm 0.0006$			&$54547.1347 \pm 0.0753$	&$54547.1180 \pm 0.1974$			\\
$e$								&0												&\phn 0.0	 assumed						&$\;\;\;0.0023 \pm 0.0002$~\tablenotemark{a}	&$0.0023 \pm 0.0006$	\\
$\omega_1$ (deg)		&\nodata										&\nodata											&$139.01 \pm 5.14\;\;$			&$137.86 \pm 13.48$					\\
$K_1$ (km/s)				&$35.8 \pm 0.3$							&$36.326 \pm 0.029$						&$36.293 \pm 0.008$				&$36.292 \pm 0.020$					\\
$K_2$ (km/s)				&$45.2 \pm 0.3$							&$44.699 \pm 0.039$						&$44.717 \pm 0.014$				&$44.718 \pm 0.038$					\\
$V_0$ (km/s)				&$-14.1 \pm 0.2\;\;\:$				&$-13.975 \pm 0.018\;\;\:$			&$-13.497 \pm 0.006\;\;\:$	&$-13.501 \pm 0.016\;\;\:$		\\
$N_{\rm obs}$				&27												&82													&16											&16				\\
$\chi^2$ primary			&\nodata										&\nodata											&92.10										&15.42			\\
$\chi^2$ secondary		&\nodata										&\nodata											&138.65									&19.42			\\
$\sigma_{\rm RV}$ primary (km/s)		&\nodata 				&0.19 (unit weight)							&0.061										&0.059			\\
$\sigma_{\rm RV}$ secondary (km/s)	&\nodata 				&\nodata											&0.126										&0.126			\\ \\
$M_1 \sin^3 i\ (M_\odot)$	&$0.163 \pm 0.003$			&$0.16090 \pm 0.00032$				&$0.16054 \pm 0.00011$ 		&$0.16054 \pm 0.00030$		\\		
$M_2 \sin^3 i\ (M_\odot)$	&$0.129 \pm 0.002$			&$0.13076 \pm 0.00024$				&$0.13030 \pm 0.00007$		&$0.13029 \pm 0.00018$		\\		
\enddata
\tablenotetext{a}{\ but see text regarding the validity of this nonzero eccentricity}
\end{deluxetable}


\subsection{12 Bo\"otis}

\begin{figure}[ht]
\centering
\includegraphics[scale=0.7]{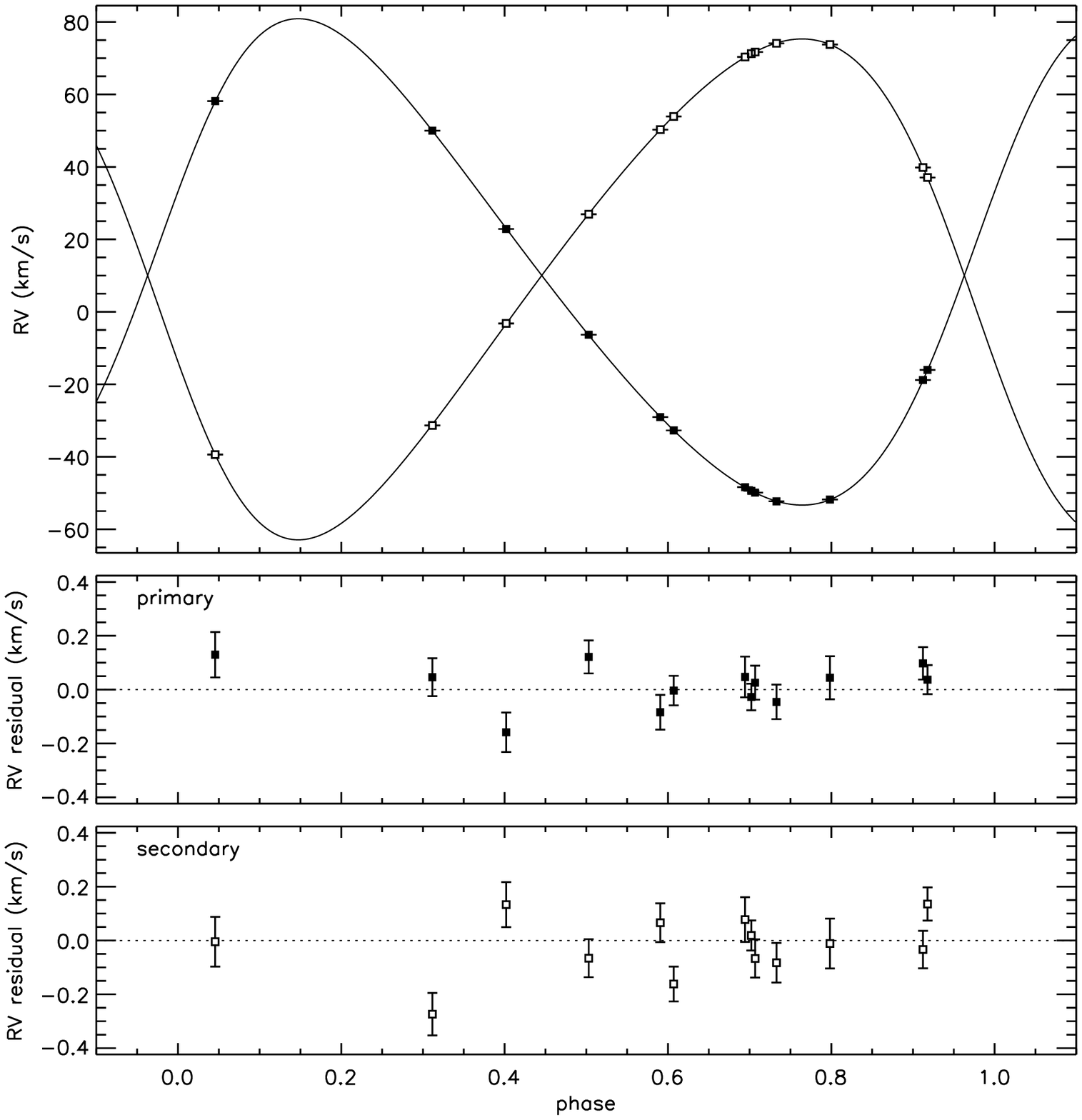}
\caption{Radial velocity measurements of the SB2 system 12 Bo\"otis.\label{12boo-fig}}
\end{figure}

The spectroscopic binary 12 Bo\"otis (HR 5304, HD 123999, HIP 69226) has received recent attention from both \citet{bod05}, who combined spectroscopic and astrometric data, and \citet{tom06}, who performed a high-precision spectroscopy-only assessment of the orbit. Orbital parameters are shown in Table~\ref{12boo-table}, and our RV data are plotted in Figure~\ref{12boo-fig}. The derived quantities for $e$ and $\omega_1$ are in excellent agreement among all three studies. Our $K_1$ and $K_2$ values, on the other hand, are smaller than those of Boden et al. and Tomkin \& Fekel by several standard deviations, and our derived systemic velocity is different as well. The discrepancy in $V_0$ may simply be ascribed to a different RV zero point, but the difference in $K_{1/2}$ deserves further scrutiny. Due to the premature conclusion of our observing program, our RV data do not fully cover the region of maximum absolute velocities around phase~$=0.15$, so the velocity amplitudes are not as reliably constrained as they might be. When dFTS observations resume, 12~Boo will be one of our highest-priority targets, so that this issue can be addressed.

Given smaller $K$ amplitudes than prior publications, we derive smaller masses as well. Using $i = 107.990^\circ \pm 0.077^\circ$ from Boden et al., we determine $M_1 = 1.4013 \pm 0.0025\ M_\odot$ and $M_2 = 1.3607 \pm 0.0024\ M_\odot$, for a relative statistical uncertainty of 0.18\%. These values are several $\sigma$ smaller than the masses derived by Boden et al. and Tomkin \& Fekel.

\begin{deluxetable}{lcccc}
\tabletypesize{\scriptsize}
\tablewidth{0pt}
\tablecaption{Orbital parameters and stellar mass estimates for 12 Bo\"otis.\label{12boo-table}}
\tablehead{\colhead{parameter} & \colhead{\citet{bod05}\tablenotemark{a}} & \colhead{\citet{tom06}} & \colhead{this work (formal RV errors)} & \colhead{this work (scaled RV errors)}}
\startdata
$P$ (days)						&$9.6045492 \pm 0.0000076$		&$9.6045529 \pm 0.0000048$ 		&adopted from T\&F					&adopted from T\&F					\\
$T$ (reduced HJD)		&$51237.7729 \pm 0.0051$			&$52400.4292 \pm 0.0035$			&$54542.2431 \pm 0.0031$	&$54542.2424 \pm 0.0042$	\\
$e$								&$0.19233 \pm 0.00086$				&$0.19268 \pm 0.00042$				&$0.1928 \pm 0.0003$			&$0.1928 \pm 0.0004$			\\
$\omega_1$ (deg)		&$286.67 \pm 0.19$						&$286.87 \pm 0.14$						&$286.79 \pm 0.12$				&$286.78 \pm 0.17$				\\
$K_1$ (km/s)				&$67.302 \pm 0.087$						&$67.286 \pm 0.037$						&$67.107 \pm 0.035$				&$67.113 \pm 0.047$				\\
$K_2$ (km/s)				&$69.36 \pm 0.10$							&$69.30 \pm 0.05$							&$69.110 \pm 	0.037$				&$69.102 \pm 0.054$				\\
$V_0$ (km/s)				&$\phn 9.551 \pm 0.051$				&$\phn 9.578 \pm 0.022$				&$10.040 \pm 0.018$				&$10.046 \pm 0.025$				\\
$N_{\rm obs}$				&49													&24													&13											&13				\\
$\chi^2$ primary			&$\sim 49.0$									&\nodata											&17.87										&10.16			\\
$\chi^2$ secondary		&$\sim 49.0$									&\nodata											&30.79										&14.02			\\
$\sigma_{\rm RV}$ primary (km/s)		&0.47						&0.11												&0.082										&0.077			\\
$\sigma_{\rm RV}$ secondary (km/s)	&0.54						&\nodata											&0.115										&0.119			\\ \\
$M_1\ (M_\odot)$		&$1.4160 \pm 0.0049$					&$1.416 \pm 0.003$						&$1.4013 \pm 0.0025$			&$1.4011 \pm 0.0031$			\\		
$M_2\ (M_\odot)$		&$1.3740 \pm 0.0045$					&$1.375 \pm 0.002$						&$1.3607 \pm 0.0024$			&$1.3608 \pm 0.0028$			\\		
\enddata
\tablenotetext{a}{\ combined fit to RV and astrometric data}
\end{deluxetable}


\subsection{V1143 Cygni}

\begin{figure}[ht]
\centering
\includegraphics[scale=0.7]{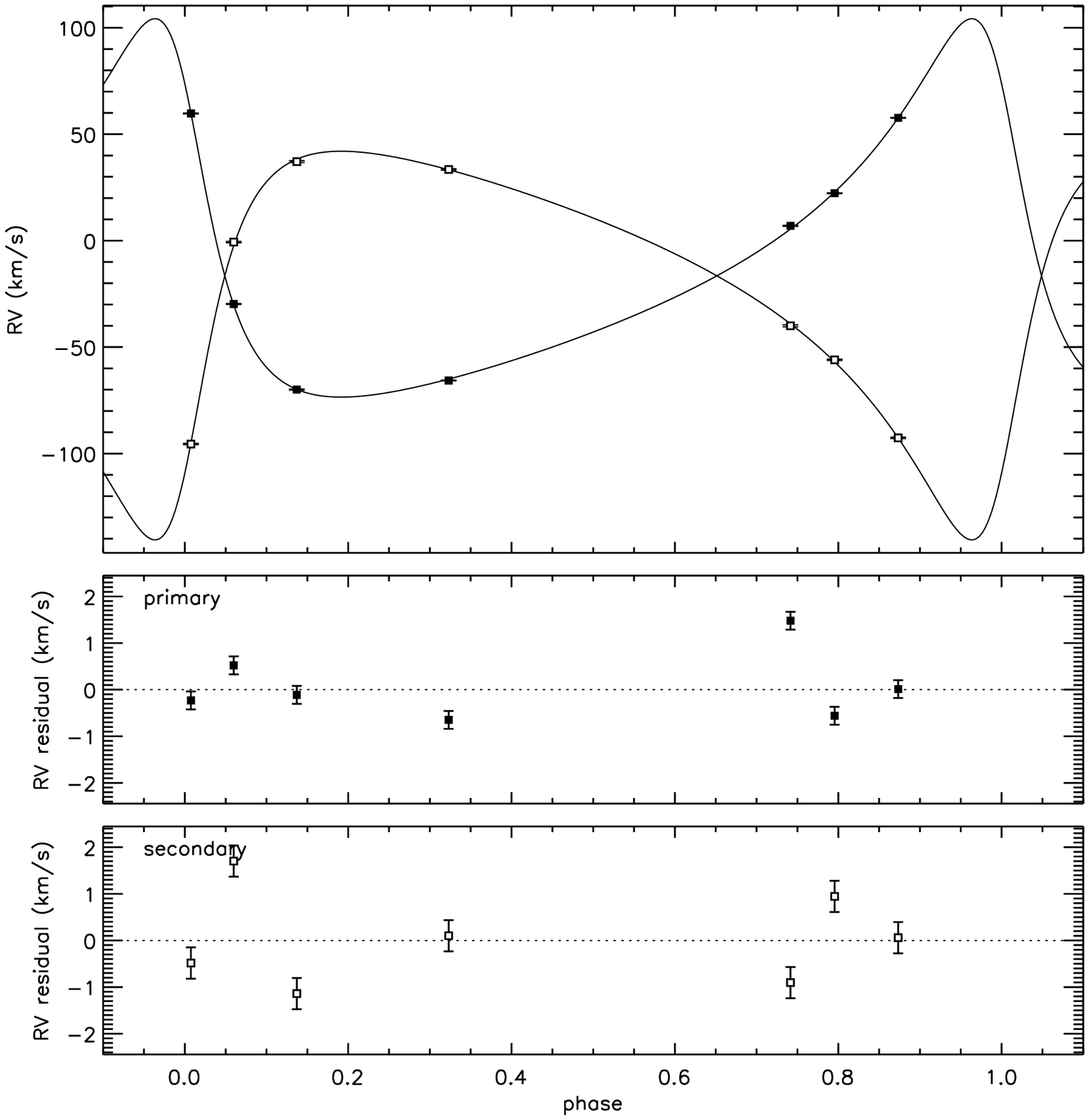}
\caption{Radial velocity measurements of the SB2 system V1143 Cygni.\label{v1143cyg-fig}}
\end{figure}

The eclipsing SB2 system of V1143 Cygni (HR 7484, HD 185912, HIP 96620) was previously analyzed by \citet{and87}. Their orbital parameters are compared to ours in Table~\ref{v1143cyg-table}, and our RV data are plotted in Figure~\ref{v1143cyg-fig}. There is broad agreement between the two sets of orbital elements, although the error bars that we derived for $K_1$ and $K_2$ are relatively large, partly because of the small number of observations, partly because the individual RV measurements had larger error bars due to larger rotational broadening of the absorption lines and lower signal-to-noise ratio. Of particular interest is the comparison of the periastron angle $\omega_1$. From precise photometric timing of the system's eclipses, \citet{gim85} detected apsidal motion (precession of the periastron point) with a period of 10,750 years, which would imply a change in the value of $\omega_1$ of $0.76^\circ$ during the $\sim 22.6$ years that elapsed between their last observations (October 1985) and our first observations (May 2008). The actual measured change in $\omega_1$ is $+0.39^\circ \pm 0.25^\circ$. The predicted periastron precession is therefore not ruled out, but is not solidly confirmed either. With more extensive observations of  V1143~Cyg, we hope to place more useful constraints on the magnitudes of the classical gravitational quadrupole and general relativity effects which cause the precession.

\citet{and87} adopt an inclination angle of $87.0^\circ \pm 1^\circ$ based upon eclipse photometry by \citet{woo71}, \citet{pop81}, and \citet{van84}. Using that same value for $i$, we estimate $M_1 = 1.3815 \pm 0.0114\ M_\odot$ and $M_2 = 1.3451 \pm 0.0100\ M_\odot$. The error bars on $K_1$ and $K_2$ dominate the error budget for the masses, so further high-accuracy spectroscopic observations of this system are clearly called for.

\begin{deluxetable}{lccc}
\tabletypesize{\footnotesize}
\tablewidth{0pt}
\tablecaption{Orbital parameters and stellar mass estimates for V1143 Cygni.\label{v1143cyg-table}}
\tablehead{\colhead{parameter} & \colhead{\citet{and87}}  & \colhead{this work (formal RV errors)}  & \colhead{this work (scaled RV errors)}}
\startdata
$P$ (days)						&$7.64075217 \pm 0.00000051$		&adopted from Andersen						&adopted from Andersen				\\
$T$ (reduced HJD)		&$42212.76652 \pm 0.00015$			&$54598.1853 \pm 0.0027$				&$54598.1835 \pm 0.0088$			\\
$e$								&$\;\;0.540 \pm 0.003$						&$\;\;0.5469 \pm 0.0010$					&$0.5484 \pm 0.0032$					\\
$\omega_1$ (deg)		&$\phn 48.6 \pm 0.02$						&$48.99 \pm 0.25$								&$48.84 \pm 0.83$						\\
$K_1$ (km/s)				&$88.20 \pm 0.20$								&$88.867 \pm 0.248$							&$89.055 \pm 0.836$					\\
$K_2$ (km/s)				&$91.10 \pm 0.40$								&$91.267 \pm 0.311$							&$91.508 \pm 0.941$					\\
$V_0$ (km/s)				&$-16.5 \pm 0.7\phn\;$						&$-16.505 \pm 0.074\phn\;$				&$-16.461 \pm 0.235\phn\;$		\\
$N_{\rm obs}$				&62														&7														&7												\\
$\chi^2$ primary			&\nodata												&63.01													&5.98											\\
$\chi^2$ secondary		&\nodata												&46.49													&5.04											\\
$\sigma_{\rm RV}$ primary (km/s)		&1.1 							&0.734													&0.789											\\
$\sigma_{\rm RV}$ secondary (km/s)	&2.2 							&1.012													&0.951											\\ \\
$M_1\ (M_\odot)$		&$1.391 \pm 0.016$							&$1.3815 \pm 0.0114$						&$1.3868 \pm 0.0334$				\\			
$M_2\ (M_\odot)$		&$1.347 \pm 0.013$							&$1.3451 \pm 0.0100$						&$1.3496 \pm 0.0308$				\\			
\enddata
\end{deluxetable}


\subsection{$\beta$ Aurigae}

\begin{figure}[ht]
\centering
\includegraphics[scale=0.7]{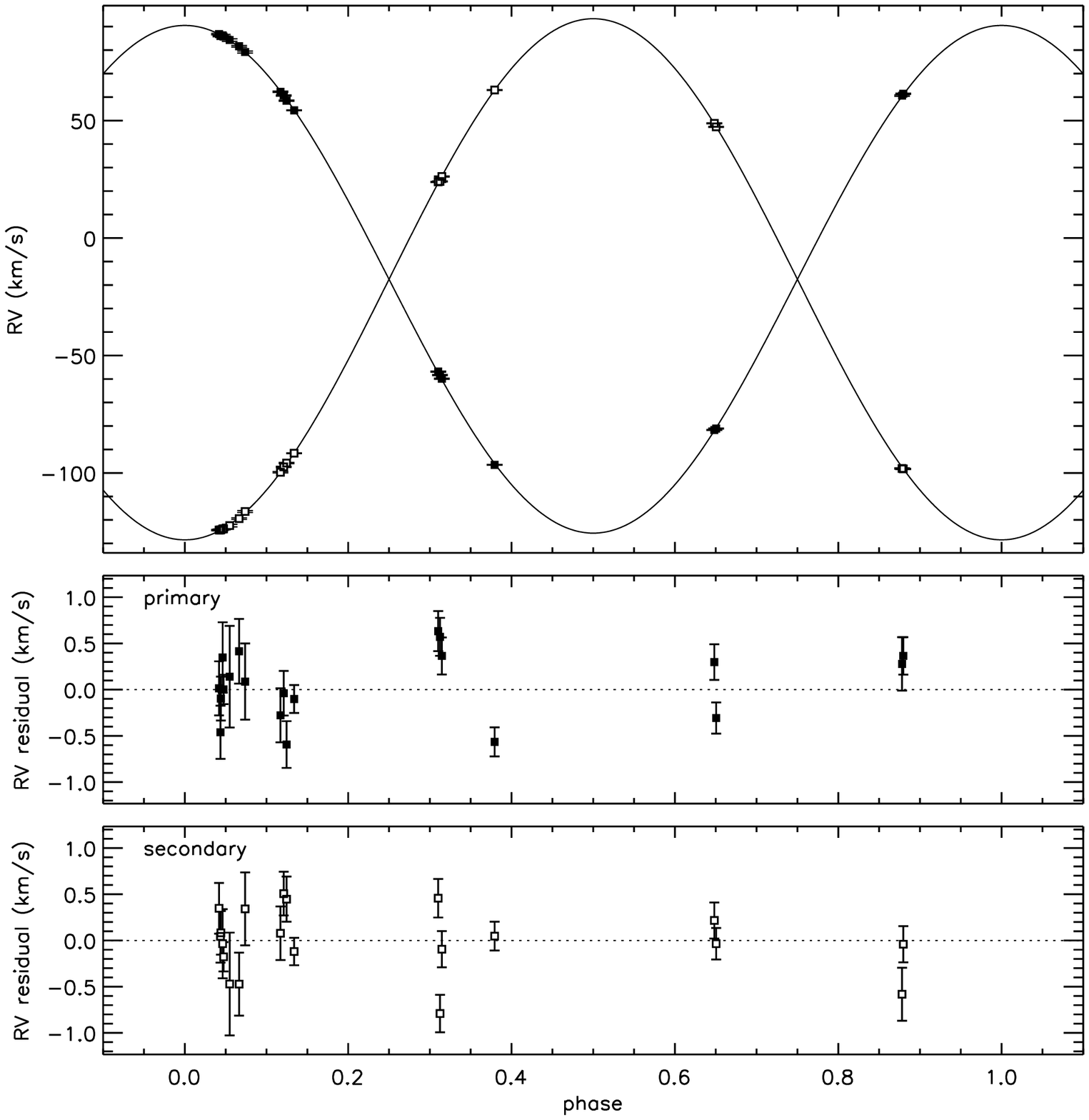}
\caption{Radial velocity measurements of the SB2 system $\beta$ Aurigae.\label{betaur-fig}}
\end{figure}

$\beta$ Aurigae (HR 2088, HD 40183, HIP 28360) is another eclipsing double-lined binary, consisting of two A2 subgiants in a close 4-day orbit. \citet{smi48} measured the RV curves of both components, and \citet{pou00} reanalyzed these data in conjunction with interferometric astrometry data from \citet{hum95} to refine the orbital parameters. Our RV curve is displayed in Figure~\ref{betaur-fig}. The phase coverage was insufficient to constrain $e$ or $\omega_1$, so we assumed a circular orbit. Our derived parameters differ from the prior two analyses (Table~\ref{betaur-table}), with $K_1$ and $K_2$ semi-amplitude values intermediate between those of Smith and those of Pourbaix. As with V1143~Cyg, this system exhibits rotational line broadening of 30--40~km/s, which increases the uncertainty of each RV measurements and thus the derived orbital parameters.

Adopting $i = 76.0^\circ \pm 0.4^\circ$ from \citet{hum95}, we calculate $M_1 = 2.3885 \pm 0.0134\ M_\odot$ and $M_2 = 2.3270 \pm 0.0130\ M_\odot$. (Pourbaix uses $i = 75.0^\circ \pm 0.73^\circ$; the source of this value is unclear.) The majority of the 0.54\% relative error in our mass values is due to the inclination angle uncertainty, so this system would be a prime follow-up target for further long-baseline spatial interferometry.

\begin{deluxetable}{lcccc}
\tabletypesize{\scriptsize}
\tablewidth{0pt}
\tablecaption{Orbital parameters and stellar mass estimates for $\beta$ Aurigae.\label{betaur-table}}
\tablehead{\colhead{parameter} & \colhead{\citet{smi48}}  & \colhead{\citet{pou00}\tablenotemark{a}} & \colhead{this work (formal RV errors)} & \colhead{this work (scaled RV errors)}}
\startdata
$P$ (days)						&$3.9600421 \pm 0.0000013$	&$3.96004 \pm 0.00000267$						&adopted from Pourbaix				&adopted from Pourbaix				\\
$T$ (reduced HJD)		&31076.719									&43915.7														&$54539.0162 \pm 0.0003$		&$54537.0362 \pm 0.0004$		\\
$e$								&0.0												&$2.75266 \times 10^{-6} \pm 0.007$		&$0.0$ assumed							&$0.0$ assumed							\\
$\omega_1$ (deg)		&0.0												&$139.043 \pm 360.0$								&\nodata 										&\nodata										\\
$K_1$ (km/s)				&$107.46 \pm 0.39$					&$110.246 \pm 1\phn$								&$108.053 \pm 0.072$				&$108.053 \pm 0.099$		\\
$K_2$ (km/s)				&$111.49 \pm 0.37$					&$\phn\;110.52 \pm 2.1$							&$110.911 \pm 0.071$				&$110.911 \pm 0.098$		\\
$V_0$ (km/s)				&$-17.06 \pm 0.27\:$					&$-15.7536 \pm 0.62$								&$-17.552 \pm 0.037\:$				&$-17.552 \pm 0.052$		\\
$N_{\rm obs}$				&21												&21																&20												&20										\\
$\chi^2$ primary			&\nodata										&\nodata														&54.19											&28.30									\\
$\chi^2$ secondary		&\nodata										&\nodata														&40.83											&21.15									\\
$\sigma_{\rm RV}$ primary (km/s)		&\nodata 				&2.740															&0.363											&0.363									\\
$\sigma_{\rm RV}$ secondary (km/s)	&\nodata 				&6.369															&0.358											&0.358									\\ \\
$M_1\ (M_\odot)$		&\nodata										&$2.4 \pm 0.1\phn$									&$2.3885 \pm 0.0129$				&$2.3885 \pm 0.0134$		\\		
$M_2\ (M_\odot)$		&\nodata										&$2.44 \pm 0.073$										&$2.3270 \pm 0.0126$				&$2.3270 \pm 0.0130$		\\		
\enddata
\tablenotetext{a}{These orbital parameters are listed in the downloadable data table at the SB9 website (http://sb9.astro.ulb.ac.be/mainform.cgi) but cannot be accessed directly from the web interface.}
\end{deluxetable}


\subsection{Mizar A}

\begin{figure}[ht]
\centering
\includegraphics[scale=0.7]{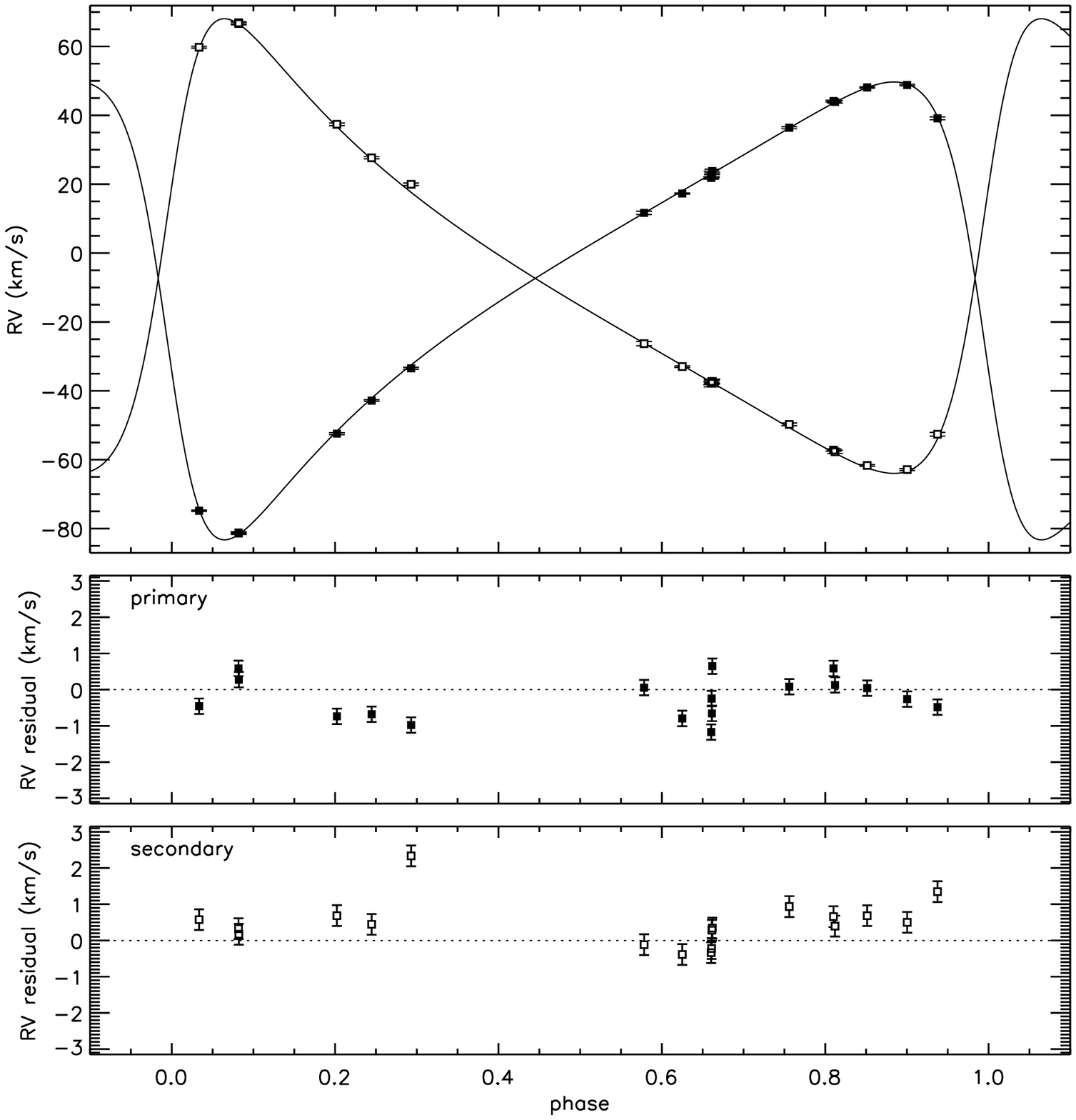}
\caption{Radial velocity measurements of the SB2 system Mizar.\label{mizar-fig}}
\end{figure}

The brighter component of the visual binary Mizar (HR 5054, HD 116656, HIP 65378) is itself a spectroscopic binary, with two early-A dwarfs in an elliptical 20-day orbit. Table~\ref{mizar-table} lists the orbital parameters measured by \citet{feh61} and the subsequent revisions computed by \citet{pou00} with astrometry data from \citet{hum98}, along with the values computed from our RV data (Figure~\ref{mizar-fig}). We find smaller velocity amplitudes and a slightly larger eccentricity for this binary system than prior researchers. Like the prior two targets, Mizar A's component spectra are moderately rotationally broadened, reducing the quality of the RV measurements.

According to Hummel et al., $i = 60.5^\circ \pm 0.3^\circ$ for this binary. Using this value, we determine that $M_1 = 2.2224 \pm 0.0221\ M_\odot$ and $M_2 = 2.2381 \pm 0.0219\ M_\odot$, for a relative mass error of 1.00\% and 0.98\% respectively. The uncertainty in $i$ is responsible for most of the mass error; if $\sigma_i$ could be reduced to $0.05^\circ$ and the errors on $K_1$ and $K_2$ can be cut in half, then the mass uncertainty would drop below 0.25\%.  

\begin{deluxetable}{lcccc}
\tabletypesize{\scriptsize}
\tablewidth{0pt}
\tablecaption{Orbital parameters and stellar mass estimates for Mizar A.\label{mizar-table}}
\tablehead{\colhead{parameter} & \colhead{\citet{feh61}} & \colhead{\citet{pou00}\tablenotemark{a}} & \colhead{this work (formal RV errors)} & \colhead{this work (scaled RV errors)}}
\startdata
$P$ (days)						&20.5386									&$20.5385 \pm 0.00013514$		&adopted from Pourbaix				&adopted from Pourbaix			\\
$T$ (reduced HJD)		&$36997.212 \pm 0.022$		&$38085.7 \pm 0.0269224$		&$54536.9882 \pm 0.0068$		&$54536.9904 \pm 0.0106$		\\
$e$								&$0.537 \pm 0.004$				&$0.529404 \pm 0.0052$			&$0.5415 \pm 0.0010$				&$0.5415 \pm 0.0016$				\\
$\omega_1$ (deg)		&$104.16 \pm 1.15$				&$\phn\;\,105.5 \pm 0.79$			&$105.21 \pm 	0.14\;\;$				&$105.27 \pm 0.23\;\;$			\\
$K_1$ (km/s)				&$68.80 \pm 0.79$					&$67.2586 \pm 0.96$					&$66.479 \pm 0.095$ 					&$66.478 \pm 0.153$				\\
$K_2$ (km/s)				&$67.60 \pm 0.91$					&$69.1796 \pm 0.77$					&$66.012 \pm 0.118$					&$66.019 \pm 0.177$				\\
$V_0$ (km/s)				&$-5.64 \pm 0.15$					&$-6.3077 \pm 0.38\;$				&$-7.342 \pm 0.052\;$				&$-7.309 \pm 0.081\;$				\\
$N_{\rm obs}$				&15											&15												&18												&18											\\
$\chi^2$ primary			&\nodata									&\nodata										&82.55											&33.70										\\
$\chi^2$ secondary		&\nodata									&\nodata										&81.44											&34.01										\\
$\sigma_{\rm RV}$ primary (km/s)		&1.87 				&1.88094										&0.556											&0.566										\\
$\sigma_{\rm RV}$ secondary (km/s)	&1.32 				&2.39922										&0.641											&0.638										\\ \\
$M_1\ (M_\odot)$		&\nodata									&$2.5 \pm 0.11$							&$2.2224 \pm 0.0221$				&$2.2228 \pm 0.0250$			\\			
$M_2\ (M_\odot)$		&\nodata									&$2.5 \pm 0.12$							&$2.2381 \pm 0.0219$				&$2.2383 \pm 0.0246$			\\			
\enddata
\tablenotetext{a}{\ with some values from the SB9 catalog \citep{pou04}}
\end{deluxetable}


\section{Results on triple systems}

\subsection{$\kappa$ Pegasi}

\begin{figure}[ht]
\centering
\includegraphics[scale=0.7]{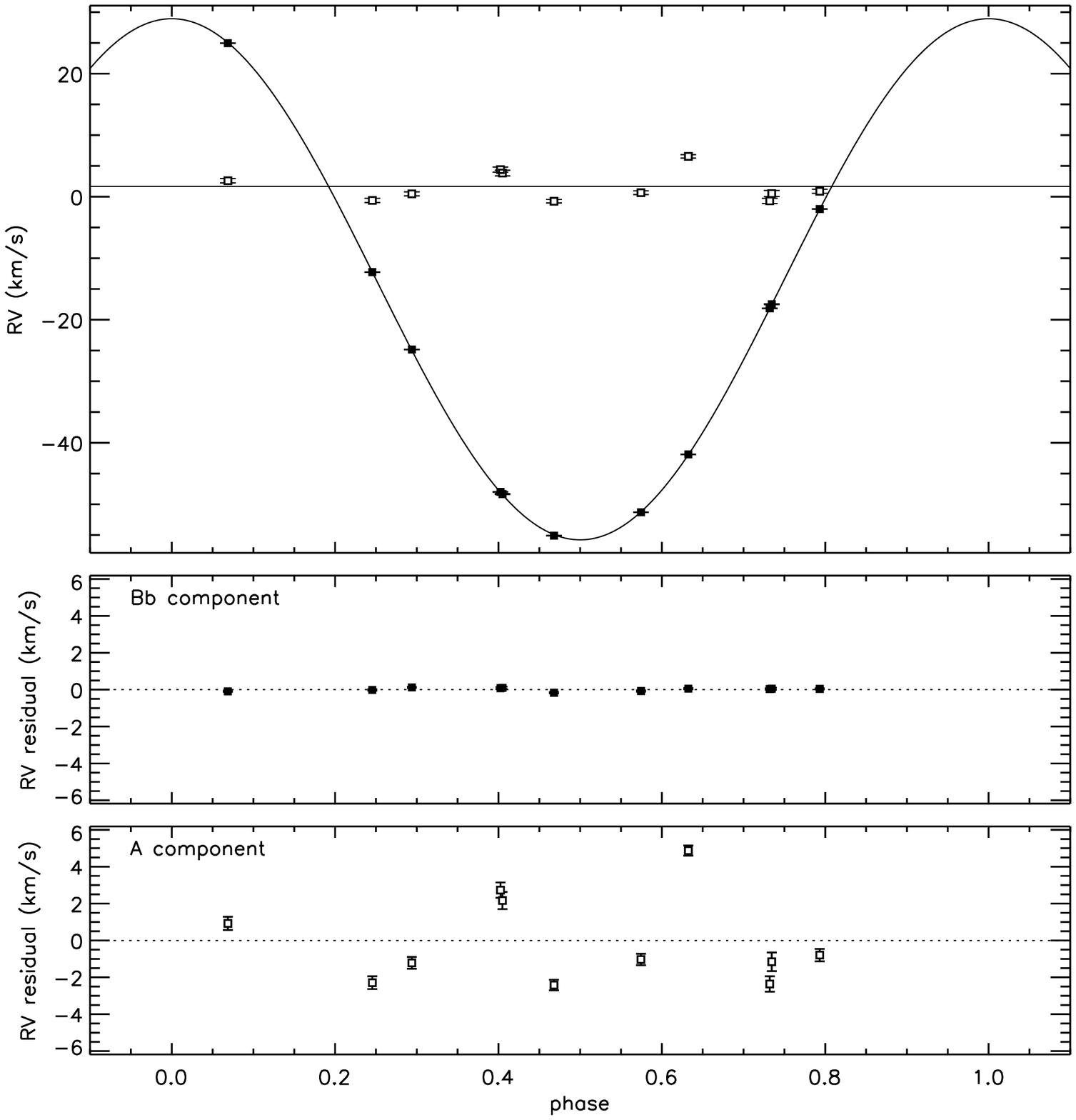}
\caption{Radial velocity measurements of the double-lined triple system $\kappa$ Pegasi. Filled symbols denote the primary of the short-period pair (``Ba''), while open symbols denote the ``A'' component of the long-period orbit.\label{kappeg-fig}}
\end{figure}

The $\kappa$ Pegasi system (HR 8315, HD 206901, HIP 107354) is a hierarchical triple, with two bright components (A \& B) in a 11.5-year orbit, and a fainter unseen component in a 6-day orbit around the B component. The canonical published orbit for this system comes from \citet{may87}, with more recent observations by \citet{kon05} (with orbit analyses published in \citet{mut06} and \citet{mut08}) and our dFTS1 prototype \citep{haj07}. Table~\ref{kappeg-table} displays the orbital parameters as measured by Mayor \& Mazeh, dFTS1, and dFTS2. Following Mayor \& Mazeh, we assume that $e=0$ for the short-period orbit. The dFTS2 observations of this system were made during two observing runs separated by $\sim 0.64$~years, so we expect the $V_0$ value of the short-period binary to change due to the long-period orbit. To account for this change, we treated the data from the two observing runs completely separately, and the results are given in two separate columns in the table. For the RV plot in Figure~\ref{kappeg-fig}, we shifted Bb component's RV values from the first observing run by $-1.198$~km/s so that the two different epochs would share the same $V_0$.

With only four RV measurements in the second observing run, the value for the $K_1$ amplitude should be considered provisional, but the change in $V_0$ is quite clear. We see an even larger change in $V_0$ as compared to the mid-1981 observations of Mayor \& Mazeh, although differences in RV zero point must be considered. Unfortunately, our prior dFTS1 observations did not yield a value of $V_0$, because the template spectra were not referenced to an absolute wavelength standard.

RV measurements of the A component are not as precise as those for the B component, because its lines are much broader: we estimate $v \sin i = 47.9$~km/s for A and 7.1~km/s for B. Even taking this fact into account, however, we find a much larger scatter of our RV measurements for A than expected from $\chi^2$ statistics. One possible explanation for this discrepancy is that the A component is also a close binary, as proposed by \citet{bea76}. We phased our RV data to their claimed 4.77-day period, but did not find any coherent cyclic pattern in the A component velocities. (\citet{mut06} see no evidence for a fourth component either.) The binarity of A is a possibility that we might explore with future data, but for the time being, this hypothesis is not supported.

\begin{deluxetable}{lccccc}
\tabletypesize{\tiny}
\tablewidth{0pt}
\tablecaption{Orbital parameters for the single-lined B component of $\kappa$ Pegasi.\label{kappeg-table}}
\tablehead{\colhead{parameter} & \colhead{\citet{may87}} & \colhead{\citet{haj07}} & \colhead{\citet{mut08}} & \colhead{this work (Oct 2007)} & \colhead{this work (June 2008)}}
\startdata
$P$ (days)						&$5.97164 \pm 0.00006$		 	&adopted from M\&M			&$5.9714971 \pm 0.0000013$			&adopted from M\&M					&adopted from M\&M						\\
$T$ (reduced HJD)		&$44801.589 \pm 0.015$			&$53681.86 \pm 0.04$		&$52402.22 \pm 0.10$						&$54400.2125 \pm 0.0005$		&$54633.0989 \pm 0.0005$			\\
$e$								&0												&adopted from M\&M			&$0.0073 \pm 0.0013$						&adopted from M\&M					&adopted from M\&M						\\
$K_1$ (km/s)				&$42.1 \pm 0.3$							&$41.572 \pm 0.257$			&\nodata												&$42.301 \pm 0.025$					&$42.527 \pm 0.037$						\\
$V_0$ (km/s)				&$-0.8 \pm 0.2\;$						&\nodata								&$-9.40 \pm 0.22$								&$-12.135 \pm 0.016\phn\;$		&$-13.333 \pm 0.022\phn\;$			\\	
$V_A$ (km/s)				&\nodata										&\nodata								&\nodata												&$\phn 1.352 \pm 0.851$			&$\phn 2.019 \pm 1.876$				\\		
$N_{\rm obs}$				&30												&9										&30														&7												&4													\\
$\sigma_{\rm RV}$ B (km/s)		&1.1 								&0.990									&0.035													&0.063											&0.093												\\
$\sigma_{\rm RV}$ A (km/s)		&\nodata 							&\nodata								&0.250													&2.084											&3.249												\\
\enddata
\end{deluxetable}

\subsection{$\eta$~Virginis}

\begin{figure}[ht]
\centering
\includegraphics[scale=0.7]{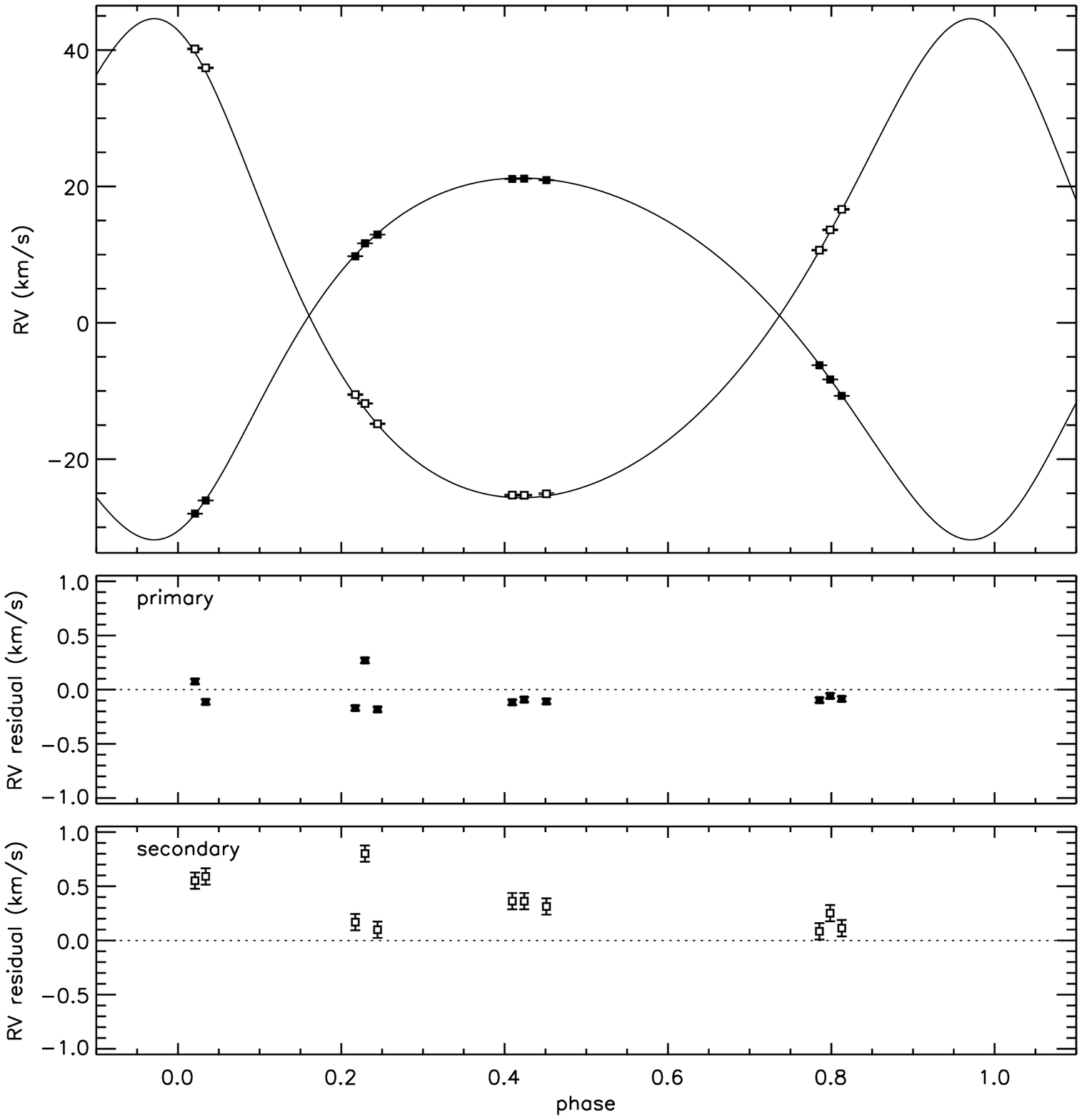}
\caption{Radial velocity measurements of the double-lined triple system $\eta$~Virginis.\label{etavir-fig}}
\end{figure}

The $\eta$~Virginis triple system (HR 4689, HD 107259, HIP 60129) was studied extensively by \citet{har92}, who combined spectroscopy and speckle interferometry to determine the orbits of both the short-period (72 day) pair and the long-period (13.1 year) grouping. \citet{hum03} made additional observations of this system with the NPOI interferometer, refining the orbital parameters and determining the inclination angle of the close binary orbit. Table~\ref{etavir-table} compares their parameters for the short-period Aa-Ab pair to our derivation. The parameters are in general agreement, except that the change in $V_0$ is larger than previously seen, and is likely due to the gravitational influence of the third component of the system. Figure~\ref{etavir-fig} shows the RV curves for the close binary, as measured by dFTS2. Note that the residuals for the secondary RV points all lie above the dotted line denoting zero residual. This offset may indicate a significant difference in the gravitational redshift or convective blueshift between the two component stars, or it may be an effect of the third component. Interestingly, Hartkopf et al. see a similar effect, but with the opposite sign, in that the $V_0$ that they derive from the Aa RV curve is $\sim 0.4$~km/s larger than $V_0$ from the Ab component.

Hartkopf et al. made a tentative spectroscopic detection of a blended Mg II 4481 feature from the faint tertiary component, which appeared to be rotationally broadened by about 160 km/s. Our instrument bandpass does not include this line, so we were unable to verify their detection, but we performed a crude three-dimensional cross-correlation using two narrow-lined template spectra plus a broad-lined A2V template. We found no evidence of a consistent RV solution for the third component.

According to the observations of \citet{hum03}, the inclination angle of $\eta$~Vir Aa-Ab is $45.5^\circ \pm 0.9^\circ$. Combining this number with our orbital parameters, we determine that $M_1 = 2.4818 \pm 0.1158\ M_\odot$ and $M_2 = 1.8745 \pm 0.0874\ M_\odot$. The inclination uncertainty is the dominant contributor to the error budget, although the velocimetry results can certainly be improved with more data points.

\begin{deluxetable}{lcccc}
\tabletypesize{\scriptsize}
\tablewidth{0pt}
\tablecaption{Orbital parameters and stellar mass estimates for $\eta$~Virginis Aa-Ab.\label{etavir-table}}
\tablehead{\colhead{parameter} & \colhead{\citet{har92}} & \colhead{\citet{hum03}} & \colhead{this work (formal RV errors)} & \colhead{this work (scaled RV errors)}}
\startdata
$P$ (days)						&$71.7919 \pm 0.0009$				&$71.7916 \pm 0.0006$		 		&adopted from Hartkopf et al.				&adopted from Hartkopf et al.	\\
$T$ (reduced HJD)		&$47583.98 \pm 0.25$				&$52321.4 \pm 0.3$					&$54403.7295 \pm 0.0938$				&$54403.6116 \pm 0.3406$	\\
$e$								&$0.272 \pm 0.009$ (Aa) 			&$0.244 \pm 0.007$					&$0.2518 \pm 0.0011$						&$0.2519 \pm 0.0040$			\\
									&$0.258 \pm 0.012$ (Ab)			&													&															&\\
$\omega_1$ (deg)		&$200.9 \pm 1.5\phn$				&$196.9 \pm 1.8$						&$197.96 \pm 0.48\phn$					&$197.21 \pm 1.74$				\\
$K_1$ (km/s)				&$26.67 \pm 0.20$						&same as Hartkopf et al.				&$26.532 \pm 0.054$							&$26.606 \pm 0.198$				\\
$K_2$ (km/s)				&$35.58 \pm 0.31$						&same as Hartkopf et al.				&$35.128 \pm 0.081$							&$35.236 \pm 0.273$				\\
$V_0$ (km/s)				&$5.24 \pm 0.19$ (Aa) 	 			&$4.9 \pm 0.2$							&$\phn 1.055 \pm 0.009$					&$1.118 \pm 0.033$				\\
									&$4.85 \pm 0.32$ (Ab)				&													&															&\\
$N_{\rm obs}$				&50												&same as Hartkopf et al.				&11														&11											\\
$\chi^2$ primary			&\nodata										&\nodata										&292.62												&25.75										\\
$\chi^2$ secondary		&\nodata										&\nodata										&263.24												&28.37										\\
$\sigma_{\rm RV}$ primary (km/s)		&1.96					&same as Hartkopf et al.				&0.129													&0.150										\\
$\sigma_{\rm RV}$ secondary (km/s)	&4.12	 				&same as Hartkopf et al.				&0.231													&0.218										\\ \\
$M_1\ (M_\odot)$		&$2.34 \pm 0.2\;$						&$2.68 \pm 0.15$						&$2.4818 \pm 0.1158$						&$2.5039 \pm 0.1246$			\\			
$M_2\ (M_\odot)$		&$1.95 \pm 0.2$~\tablenotemark{a}			&$2.04 \pm 0.10$		&$1.8745 \pm 0.0874$						&$1.8907 \pm 0.0932$			\\
\enddata
\tablenotetext{a}{\ assumed}
\end{deluxetable}


\section{Summary}

Our results demonstrate that dFTS technology is well-suited to high-accuracy radial velocity measurements of double-lined spectroscopic systems. We have determined the orbital parameters of six binary systems, matching or improving the published values for the masses of the component stars. We also observed two double-lined triple systems, providing some constraints on the nature of their stars.

For our future observational programs for spectroscopic binary stars, we are motivated by an assortment of specific scientific goals for which the capabilities of a dFTS are particularly applicable:

1. The most immediate goal is to continue to improve the accuracy of orbital parameters of binary systems, particularly the $K$ amplitudes, and thus measure stellar masses more accurately. These advancements in spectroscopic capabilities must proceed in parallel with better astrometric measurements, as determined by current and future long-baseline spatial interferometers.

2. With high-accuracy RV measurements spanning longer periods of time, we will be able to detect and quantify secular changes in binaries' systemic velocities ($V_0$) due to tertiary companions. As discussed by \citet{tok06}, the presence of a tertiary companion has significant implications for the formation of close binaries.

3. Observations of near-circular binary orbits will confirm or refute small nonzero eccentricities, thus providing observational validation for theories of tidal circularization and the influence of external gravitational perturbations such as Kozai resonances.

4. For highly elliptical systems like V1143~Cyg, long-term observing programs can measure changes in periastron angle to test theories of apsidal precession due to classical and relativistic effects.

5. Because the instrumental profile of a dFTS is easy to calculate {\it a priori} from the delay sampling function, we can measure spectral line broadening very accurately, {\it e.g.} to determine the projected rotational velocities of stellar components and thus shed light on tidal spin-up/spin-down mechanisms.

\pagebreak

\acknowledgments

This research was funded in part by ARH's NSERC Discovery Grant.

We are grateful to the day crew at Steward Observatory --- Jeff Fearnow, Dave Harvey, Bob Peterson, Gary Rosenbaum, and Bill Wood --- for their assistance with the transport and installation of dFTS2, and we thank telescope operators Geno Bechetti, Dennis Means, and Peter Milne for their expertise in operating the telescope on our behalf. We also express our appreciation to the Director of the Steward Observatory for granting us telescope time over an extended period.

We are greatly indebted to the skilled instrument builders in the USNO Machine Shop --- Gary Wieder, Dave Smith, Tie Siemers, and John Evans --- for fabricating all of the custom optomechanical elements of dFTS2, as well as the thermal enclosure. We also thank the USNO Astrometry Department for travel support and salary support during the initial stages of this observing program, and thanks also go to the USNO Time Services Division for lending us packing crates for shipment of our instrument to Kitt Peak. 

This research has made use of the SIMBAD database, operated at CDS, Strasbourg, France; NASA's Astrophysics Data System; and the SB9 catalog of \citet{pou04}. Richard O. Gray is to be commended for making his SPECTRUM codes so easy to install and use. Thanks also to Farnoud Kazemzadeh for a critical reading of the manuscript.


\end{document}